\documentclass{caosp}
\usepackage{graphicx}
\usepackage{longtable}

\articleNo{000}
\pubyear{2018}
\volume{00}
\volnumber{1}
\firstpage{1}
\received{March 00, 2018}
\accepted{???}

\begin{document}

\htitle{Long-period SU UMa dwarf nova V1006 Cygni: outburst activity and variability ...}
\hauthor{E.P. Pavlenko, S.Yu. Shugarov, A.O. Simon, A.A. Sosnovskij, K.A. Antonyuk et al.}

\title{Long-period SU UMa dwarf nova V1006 Cygni: outburst activity and variability at different brightness states in 2015 -- 2017 }

\author{
       E.P.\,Pavlenko\inst{1}
       \and
       S.Yu.\,Shugarov\inst{2,3}
       \and
       A.O.\,Simon\inst{4}
       \and
       A.A.\,Sosnovskij\inst{1}
       \and
       K.A.\,Antonyuk\inst{1}
       \and
       O.I.\,Antonyuk\inst{1}
       \and
       A.V.\,Shchurova\inst{2}
       \and
       A.V.\,Baklanov\inst{1}
       \and
       Ju.V.\,Babina\inst{1}
       \and
       A.S.\,Sklyanov\inst{5}
       \and
       V.V.\,Vasylenko\inst{4}
       \and
       V.G.\,Godunova\inst{6}
       \and
       I.\,Sokolov\inst{7}
       \and
       I.V.\,Rudakov\inst{8}
       }

\institute{
         Federal State Budget Scientific Institution ``Crimean Astrophysical Observatory of RAS'', 
         Nauchny, 298409, Republic of Crimea,
         \and
         \lomnica,
         \and
         Sternberg Astronomical Institute, Moscow State University,
         Universitetskij pr., 13, Moscow, 119991, Russia,
         \and
         Astronomy and Space Physics Department, Taras Shevshenko National University of Kyiv, 
         Volodymyrska str. 60, Kyiv, 01601, Ukraine
         \and
         Kazan (Volga region) Federal University, Kazan, 420008, Kremlyovskaya 18, Russia
         \and
         ICAMER Observatory of NASU 27 Acad. Zabolotnoho str. 03143, Kyiv, Ukraine
         \and 
         Institute of Astronomy, Russian Academy of Sciences, Terskol Branch, Settlement Terskol,               Kabardino-Balkarian Republic, 361605, Russian Federation.
         \and
         Observatory of the National Scientific and Practical, Educational and Wellness Center 
         ``Bobek'', Askarov str. 47, Almaty, 050027, Republic of Kazakhstan
         }

\date{march 00, 2018}
\maketitle

\begin{abstract}
CCD photometric observations of the dwarf nova V1006 Cyg were carried out in 2015 -- 2017  with 11 telescopes located at  7 observatories. They covered the 2015 superoutburst with rebrightening, five normal outbursts of $\sim 4$-day duration and one wide outburst that lasted at least seven days.  The interval between normal outbursts was 16 and 22 days, and  between superoutbursts is expected to be longer  than 124 days.  The positive superhumps with the mean period of $0^d.10544(10)$  and   $0^d.10406(17)$ were
detected  during the 2015 superoutburst and during the short-term quiescence between rebrightening and the start of the first normal outburst, respectively.
During a wide 2017 outburst the orbital period  $0^d.09832(15)$  was found.
The amplitude of this signal was $\sim 2.5$ times larger at the outburst decline
than at its end. During the quiescence stage between the first and the
second normal outbursts in 2015 we possibly detected the negative
superhumps with the period of $0^d.09714(7)$. In all other cases of quiescence
we found only the quasi-periodic brightness variations on a time scale of 20--30 minutes with a different degree of coherence and a variable amplitude reaching 0.5 mag in extremal cases.

\end{abstract}

%%%%%%%%%%%%%%%%%%%%%%%%%%%%%%%%%%%%%%%%%%%%%%%%%%%%%%%%%%%%%%%%%%%%%%%%%%%%%
%                       S E C T I O N I N G
% Any section starts with the command \section as shown below, with the
% title in Initial Capitals and lowercase only. Do not number the sections
% - let LaTeX do that for you - and do not end them by a "." (dot).
%
% The (sub)section titles are typeset in boldface; so, if working in the
% mathematics mode in (sub)section titles, you must use \boldmath and
% enclose it into curly brackets, e.g. "{\bolmath $R^{2}$}".
%%%%%%%%%%%%%%%%%%%%%%%%%%%%%%%%%%%%%%%%%%%%%%%%%%%%%%%%%%%%%%%%%%%%%%%%%%%%%
%\section{Introduction}
%%%%%%%%%%%%%%%%%%%%%%%%%%%%%%%%%%%%%%%%%%%%%%%%%%%%%%%%%%%%%%%%%%%%%%%%%%%%%
%                       L A B E L
% The label command is very convenient for you when referring to sections,
% subsections,..., tables, figures as well as to equations (see commands
% \ref and \pageref). In the case of figure and/or table environments the
% \label command should always be put after the \caption command to
% preserve proper numbering. When using the \label command the file must
% be compiled twice to get proper cross-references.
%%%%%%%%%%%%%%%%%%%%%%%%%%%%%%%%%%%%%%%%%%%%%%%%%%%%%%%%%%%%%%%%%%%%%%%%%%%%%
\label{intr}

\section{Introduction}

Dwarf novae are a  subclass of cataclysmic variables. They consist of a pair of dwarfs. The red (or brown) 
dwarf fills in its Roshe Lobe and loses  material onto the white dwarf, creating an accretion disk around it 
(Warner 1995). The matter in disk accumulates until reaching some critical conditions that are necessary for 
transition of the disk from the cool  and neutral one to the hot and ionized one. This thermal disk instability 
(Osaki 1996; Hoshi 1979; Meyer and Meyer-Hofmeister 1981 ) leads to the dwarf nova outburst. SU UMa-type dwarf 
novae are a special subclass of the dwarf novae  possessing  two types of outbursts -- normal outbursts and 
superoutbursts.  Several  normal outbursts that have typical duration of  2-5 days and an outburst amplitude of 
2 -- 3 mag are placed between superoutbursts. The last ones have a higher amplitude and longer duration (Warner 1995).  
The time between neighbor superoutbursts (supercycle)  is  a more-or- less stable value and could be  a
characteristic of the dwarf nova. Thus  stars with a supercycle  $<100^d$  are called ``active'' dwarf novae,  
these with shortest (19 -- 48 days) supercycles  are ER UMa-type dwarf novae (Kato et al. 2013).  
 WZ-type dwarf novae have the longest supercycles lasted  years -- decades (Kato 2015).  
There could be several cycles (time between normal outbursts)  during the supercycle.  Both types of 
outbursts are considered to be a result of the combination of thermal and tidal instabilities (Osaki 1989; Osaki 1996). 

Only during the outbursts the periodic brightness variations (so-called ``positive superhumps'') with a period of a few percent longer than the orbital one are observed. According to a modern point of view, these superhumps 
originate from a precessing eccentric disk, and eccentricity that  is believed to be caused by the 3:1 
resonance in the disk (Whitehurst 1988; Hirose and Osaki 1990; Lubow  1991; Wood et al. 2011). The requirement for the noted resonance is the mass ratio of the components $m_2/m_1 < 0.3$ that is implemented for SU UMa stars with orbital periods between 
76 min and 3.18 hours (Knigge 2006).  These periods  include a known  ``period gap''  between 2 and 3 
hours for the cataclysmic variables distribution. However, as it was shown on the basis of a large  number 
of examples (Pavlenko et al. 2014; Kato et al. 2017a), SU UMa stars in the period gap display rather  
extension of  monotonic  decrease of  the stars number toward the long-periodic border  of the gap  instead of  
an abrupt shortage of them. 

In any state of the outburst activity there could be variations that are a few 
percent less than the orbital one -- so-called ``negative superhumps'' (Udalski 1988; Harvey et al. 1995; 
Ringwald et al. 2012). The origin of negative superhumps is usually considered as a result of retrograde 
precession of a tilted accretion disk (Wood, Burke 2007). The number of known negative superhumpers is 
much smaller than that of known positive superhumpers among SU UMa stars and currently comprises about 
15 binaries (Pavlenko, under preparation).

V1006 Cyg was discovered as a  dwarf nova  by Hoffmeister (1963a, 1963b) with a photographic  
range of 16--18 mag. Gessner (1966) and Bruch et al. (1987)  found it in an outburst later on. Bruch and Schimpke (1992) 
also identified this object as a dwarf nova from its spectrum. Sheets et al. (2007) used radial velocities 
to find the orbital period of $0^d.09903(9)$.  The second time V1006 Cyg  was found in rather wide outbursts 
lasting more than six days in 2007 and more than five days in 2009 (Pavlenko et al. 2014), but no superhumps were observed. Instead  the orbital period was detected in both outbursts by Pavlenko et al. (2014).

 Kato et al. (2016) informed that the start of the first genuine superoutburst  of V1006 Cyg was detected on July, 12  2015 by Shappee et al. by the All-Sky Authomated Survey for Supernovae (ASAS-SN).  The 2015 superoutburst was studied in detail  by the VSNET campaign (Kato et al. 2016).  The  $0^d.10541$ superhump period  was found that finally established  this object to be an SU UMa-type dwarf nova in the period gap, i.e, a long-periodic SU UMa dwarf nova. Recently, Kato et al. (2009) showed that the evolution of superhumps of SU UMa stars has three stages.  These stages were  also found in the period of superhump variations of V1006 Cyg (Kato et al. 2016). The authors estimated mass ratio for this binary as 0.26-0.33 solar masses. However, the cycle, supercycle of V1006 Cyg and its behavior in quiescence was unknown. This motivated us  to undertake the next multi-site campaign  on a long-term study  of this dwarf nova  immediately at the end of  the VSNET campaign.
\vspace{-1mm}

\section{Observations and data reduction}

Photometric investigations of V1006 Cyg  have been carried out with 11 telescopes located  at 7 observatories 
in photometric system close to the Johnson - Cousins $UBVR_CI_C$ system or in the integral light (symbol "C"), 
see  Table~1: Journal of observations. 
In this paper we analyse only $BVR_C$ data.
%\newpage
%\input{Tab_1.tab}
%\vspace{5mm}
The standard data reduction included a flat-fielding, bias and dark signal removal. We used the MAXIM DL  
and V. Goranskij (http://www.vgoranskij.net/software/) WinFit packages  to measure the brightness of variable,  
comparison and check stars.
%\newpage
%\input{Tab_1.tab}
%\vspace{5mm}  
The data were referred to AAVSO comparison star No~140.  Its brightness 
was measured relatively  to the known magnitudes in the vicinity of CH Cyg (Henden and Munari, 2006). 
For this star we obtained $B=14^m.89, V=13^m.99, R_C=13^m.47, I_C=13^m.03$. The corresponding  AAVSO data 
are  $B=14^m.96, V=13^m.96, R_C=13^m.45, I_C=12^m.97$. For the analysis we used our data. We combined  the 
data from different  locations by  adding  the corresponding corrections which we calculated for common 
nights of observations  for different telescopes. 
The intrinsic accuracy was measured as a standard deviation 
for the number of check stars in regard to the comparison one. It depended on the telescope, weather condition, 
exposure time  and the brightness of object. At the maximum it was $0^m.005 - 0^m.05$ and at minima $0^m.03 - 0^m.1$. 

\newpage
\begin{center}
\footnotesize
\tabcolsep 5.5pt
\begin{longtable}[pth]{llrlllll}
\\[-10mm]
\caption{
Journal ob observations.
JD~=~JD$^*$ + 2\,450\,000.}\\
\hline\hline
Date&JD$^*$start-end&N&Tel.&Obs.&CCD&Band&\hspace{-4mm}Stage\\
\hline
\endfirsthead
\caption{Continued.}\\
\hline
Day&JD$^*$start-end&N&Tel.&Obs.&CCD&Band&\hspace{-4mm}Stage\\
\hline
\endhead
\hline
\endfoot
\hline\hline
\endlastfoot
%\hline
{\bf 2015} & & & & & & & \\
17.07&7221.279-.380&87 & 2.6m&CrAO   &APOGEE E47   &Rc      &S  \\
18.07&7222.262-.564&97 & 50cm&SAI    &APOGEE Al.U16&CUBVRcIc&S  \\
19.07&7223.291-.564&92 & 50cm&SAI    &APOGEE Al.U16&CUBVRcIc&S  \\
20.07&7224.225-.264&11 & 50cm&SAI    &APOGEE Al.U16&CUBVRcIc&S  \\
20.07&7224.271-.546&230& 38cm&CrAO   &APOGEE E47   &C       &S  \\
21.07&7225.225-.299&28 & 50cm&SAI    &APOGEE Al.U16&CUBVRcIc&S  \\
21.07&7225.264-.547&191& 38cm&CrAO   &APOGEE E47   &C       &S  \\
21.07&7226.224-.290&22 & 50cm&SAI    &APOGEE Al.U16&CUBVRcIc&S  \\
22.07&7226.547-.460&72 & 38cm&CrAO   &APOGEE E47   &C       &S  \\
23.07&7227.280-.552&240& 38cm&CrAO   &APOGEE E47   &C       &S  \\
24.07&7228.279-.558&186& 38cm&CrAO   &APOGEE E47   &C       &S  \\
25.07&7229.269-.563&183& 38cm&CrAO   &APOGEE E47   &C       &S  \\
26.07&7230.275-.562&126& 38cm&CrAO   &APOGEE E47   &C       &S  \\
27.07&7231.348-.564&200& 38cm&CrAO   &APOGEE E47   &C       &R  \\
28.07&7232.278-.567&197& 38cm&CrAO   &APOGEE E47   &C       &R  \\
29.07&7233.256-.564&195& 38cm&CrAO   &APOGEE E47   &C       &R  \\
30.07&7234.262-.555&198& 38cm&CrAO   &APOGEE E47   &C       &Q  \\
31.07&7235.283-.558&182& 38cm&CrAO   &APOGEE E47   &C       &Q  \\
31.07&7235.367-.548&124& 60cm&SL     &FLI ML3041   &CBVRcIc &Q  \\
01.08&7236.254-.525&124& 38cm&CrAO   &APOGEE E47   &C       &Q  \\
01.08&7236.316-.542&178& 60cm&SL     &FLI ML3041   &C       &Q  \\
02.08&7237.377-.560&71 & 38cm&CrAO   &APOGEE E47   &C       &Q  \\
02.08&7237.465-.339&143& 60cm&SL     &FLI ML3041   &C       &Q  \\
03.08&7238.288-.545&102& 38cm&CrAO   &APOGEE E47   &C       &O1 \\
03.08&7238.303-.323&17 & 18cm&SL     &SBIG ST-10XME&C       &O1 \\
04.08&7239.247-.571&177& 38cm&CrAO   &APOGEE E47   &C       &O1 \\
04.08&7239.298-.550&79 & 18cm&SL     &SBIG ST-10XME&C       &O1 \\
04.08&7239.387-.516&50 & 28cm&KFU    &QSI 583wsg   &C       &O1 \\
05.08&7240.269-.566&401& 38cm&CrAO   &APOGEE E47   &C       &Q  \\
06.08&7241.337-.499&49 & 28cm&KFU    &QSI 583wsg   &C       &Q  \\
07.08&7242.277-.504&60 & 28cm&KFU    &QSI 583wsg   &C       &Q  \\
07.08&7242.252-.558&200& 38cm&CrAO   &APOGEE E47   &C       &Q  \\
08.08&7243.289-.211&60 & 28cm&KFU    &QSI 583wsg   &C       &Q  \\
08.08&7243.254-.358&235& 2.6m&CrAO   &APOGEE E47   &C       &Q  \\
09.08&7244.377-.398&47 & 2.6m&CrAO   &APOGEE E47   &C       &Q  \\
09.08&7244.314-.592&153& 60cm&SL     &FLI ML3041   &C       &Q  \\
10.08&7245.211-.506&99 & 28cm&KFU    &QSI 583wsg   &C       &Q  \\
10.08&7245.409-.593&102& 60cm&SL     &FLI ML3041   &C       &Q  \\
11.08&7246.218-.449&80 & 28cm&KFU    &QSI 583wsg   &C       &Q  \\
11.08&7246.548-.588&23 & 60cm&SL     &FLI ML3041   &C       &Q  \\
12.08&7247.238-.351&17 & 28cm&KFU    &QSI 583wsg   &C       &Q  \\
13.08&7248.204-.501&91 & 28cm&KFU    &QSI 583wsg   &C       &Q  \\
14.08&7249.279-.307&17 & 38cm&CrAO   &APOGEE E47   &C       &Q  \\
15.08&7250.340-.375&24 & 38cm&CrAO   &APOGEE E47   &C       &Q  \\
16.08&7251.302-.333&22 & 38cm&CrAO   &APOGEE E47   &C       &Q  \\
17.08&7252.317-.356&26 & 38cm&CrAO   &APOGEE E47   &C       &O2 \\
17.08&7252.370-.378&5  &125cm&CrAO   &ProLine 23042&Rc      &O2 \\
18.08&7253.360-.362&5  &125cm&CrAO   &ProLine 23042&VRcIc   &O2 \\
18.08&7253.302-.470&113& 38cm&CrAO   &APOGEE E47   &C       &O2 \\
19.08&7254.349-.403&22 &125cm&CrAO   &ProLine 23042&VRcIc   &O2 \\
20.08&7255.318-.375&25 &125cm&CrAO   &ProLine 23042&VRcIc   &O2 \\
21.08&7256.334-.343&5  &125cm&CrAO   &ProLine 23042&VRcIc   &O2 \\
24.08&7259.260-.271&5  &125cm&CrAO   &ProLine 23042&VRcIc   &Q  \\
25.08&7260.274-.295&9  &125cm&CrAO   &ProLine 23042&VRcIc   &Q  \\
26.08&7261.316-.321&3  &125cm&CrAO   &ProLine 23042&VRcIc   &Q  \\
27.08&7262.302-.306&3  &125cm&CrAO   &ProLine 23042&Rc      &Q  \\
28.08&7263.343-.347&3  &125cm&CrAO   &ProLine 23042&VRcIc   &Q  \\
29.08&7264.266-.270&3  &125cm&CrAO   &ProLine 23042&VRcIc   &Q  \\
31.08&7266.297     &1  & 70cm&Lisnyky&FLI PL4710   &UBVRcIc &Q  \\
01.09&7267.334     &1  & 70cm&Lisnyky&FLI PL4710   &UBVRcIc &Q  \\
01.09&7267.258-.354&45 &125cm&CrAO   &ProLine 23042&Rc      &Q  \\
02.09&7268.259-.268&5  &125cm&CrAO   &ProLine 23042&Rc      &Q  \\
02.09&7268.284     &1  & 70cm&Lisnyky&FLI PL4710   &UBVRcIc &Q  \\
03.09&7269.257-.349&43 &125cm&CrAO   &ProLine 23042&Rc      &Q  \\
04.09&7270.393-.567&723& 2.6m&CrAO   &APOGEE E47   &C       &Q  \\
08.09&7274.248-.256&5  &125cm&CrAO   &ProLine 23042&Rc      &Q  \\
09.09&7275.244-.248&3  &125cm&CrAO   &ProLine 23042&Rc      &O3 \\
09.09&7275.453-.571&207& 2.6m&CrAO   &APOGEE E47   &C       &O3 \\
10.09&7276.247-.259&6  &125cm&CrAO   &ProLine 23042&Rc      &O3 \\
11.09&7277.257-.475&97 &125cm&CrAO   &ProLine 23042&VRcIc   &O3 \\
15.09&7280.248-.257&5  &125cm&CrAO   &ProLine 23042&Rc      &Q  \\
15.09&7281.228-.232&3  &125cm&CrAO   &ProLine 23042&Rc      &Q  \\
15.09&7281.277     &1  & 70cm&Lisnyky&FLI PL4710   &UBVRcIc &Q  \\
16.09&7282.368-.473&50 &125cm&CrAO   &ProLine 23042&Rc      &Q  \\
16.09&7282.297     &1  & 70cm&Lisnyky&FLI PL4710   &UBVRcIc &Q  \\
17.09&7283.400-.445&22 &125cm&CrAO   &ProLine 23042&Rc      &Q  \\
17.09&7283.260     &1  & 70cm&Lisnyky&FLI PL4710   &Rc      &Q  \\
18.09&7284.230-.476&110&125cm&CrAO   &ProLine 23042&Rc      &Q  \\
18.09&7284.250     &1  & 70cm&Lisnyky&FLI PL4710   &UBVRcIc &Q  \\
19.09&7285.334     &1  & 70cm&Lisnyky&FLI PL4710   &Rc      &Q  \\
20.09&7286.334     &1  & 70cm&Lisnyky&FLI PL4710   &UBVRcIc &Q  \\
22.09&7288.242     &1  & 70cm&Lisnyky&FLI PL4710   &UBVRcIc &Q  \\
25.09&7291.374     &1  & 70cm&Lisnyky&FLI PL4710   &UBVRcIc &Q  \\
05.10&7301.134     &1  & 60cm&Almaty &SXVR-35H     &C       &Q  \\
06.10&7302.198     &1  & 60cm&Almaty &SXVR-35H     &C       &Q  \\
07.10&7303.252     &1  & 60cm&Almaty &SXVR-35H     &C       &Q  \\
08.10&7304.240     &1  & 60cm&Almaty &SXVR-35H     &C       &Q  \\
15.10&7310.306     &1  & 70cm&Lisnyky&FLI PL4710   &UBVRcIc &Q  \\
16.10&7311.189-.340&66 & 38cm&CrAO   &APOGEE E47   &C       &Q  \\
20.10&7315.206-.28 &97 & 38cm&CrAO   &APOGEE E47   &C       &O5 \\
30.10&7325.2       &1  & 38cm&CrAO   &APOGEE E47   &C       &Q  \\
31.10&7326.277     &1  & 70cm&Lisnyky&FLI PL4710   &Rc      &Q  \\
01.11&7327.244     &1  & 70cm&Lisnyky&FLI PL4710   &Rc      &Q  \\
05.11&7332.149-.274&172&125cm&SAI    &VersAray1300 &C       &Q  \\
11.11&7338.258     &1  & 70cm&Lisnyky&FLI PL4710   &Rc      &Q  \\
24.12&7381.215     &1  & 70cm&Lisnyky&FLI PL4710   &Rc      &Q  \\
{\bf 2016} &             &   &     &       &             &        &   \\
2.04 &7481.449-.600&180& 2.6m&CrAO   &APOGEE E47   &BVRc    &Q  \\
5.05 &7514.430-.505&216& 2.6m&CrAO   &APOGEE E47   &C       &Q  \\
20.08&7621.359-.580&47 & 70cm&Lisnyky&FLI PL4710   &UBVRcIc &O6 \\
21.08&7622.289-.578&115& 70cm&Lisnyky&FLI PL4710   &UBVRcIc &O6 \\
{\bf 2017} &             &   &     &       &             &        &   \\
16.10&8043.257-.406&201& 70cm&Lisnyky&FLI PL4710   &UBVRcIc &O8W\\
16.10&8043.178-.451&376& 38cm&CrAO   &APOGEE E47   &C       &W  \\
17.10&8044.217-.393&483& 38cm&CrAO   &APOGEE E47   &C       &W  \\
18.10&8045.206-.364&214& 38cm&CrAO   &APOGEE E47   &C       &W  \\
19.10&8046.208-.359&103& 38cm&CrAO   &APOGEE E47   &C       &W  \\
25.11&8083.285-.380&124& 2m  &Terskol&FLI PL4301   &B       &Q  \\
27.11&8085.228-.373&191& 2m  &Terskol&FLI PL4301   &B       &Q  \\
%\hline\hline
\end{longtable}
\end{center}
\vspace{-7mm}
{\footnotesize 
{\bf Description of columns:}

Date: calendar data.

JD$^*$ start-end: beginning and end of the observational run. 

N - number of observations. 

Tel.: size of the telescope's objective. 

Obs. -- Observatory: CrAO - Crimean Astrophysical Observatory, Republic of Crimea; SL - Star\'a Lesn\'a 
Observatory of the Astronomical institute of the Slovak Academy of Sciences, Slovakia; 
SAI - Southern Station of the Sternberg Astronomical Institute, Crimea; 
Lisnyky - Taras Shevchenko National University of Kyiv, Ukraine; 
Almaty - Observatory of the NSPEWC Bobek, Republic of Kazakhstan;
Terskol - Terskol observatory, Russian Academy of Sciences, Terskol Branch,
Settlement Terskol, Kabardino-Balkarian Republic;
%Terskol - CAMER Observatory of the National Academy of XXX ZZZ, North-Caucasus; 
KFU - North-Caucasus Astronomical Station of Kasan Federal University. 

CCD: CCD camera type, abbreviation "APOGEE Al.U16" is APOGEE Alta U16M. 

Band: passband $UBVR_cI_c$, C-integral light.

Stage: Designations of activity types: S - superoutburst, R - rebrightening, W - wide outburst without superhumps, 
O - normal outburst, Q - quiescence}.
\vspace{2mm}

\vspace{5mm}

In the case of a low signal-to-noise ratio the data have been stacked to reach the acceptable accuracy. 
%\newpage
%\input{Tab_1.tab}
%\vspace{5mm}
We used the Stellingwerf method for the time-series analysis with Pelt package ISDA (Pelt 1980).

\section{Analysis of the 2015-17 light curves}

\subsection{Long-term light curve, cycles and supercycle}
We observed  V1006 Cyg  since its 2015 superoutburst from July 17, 2015 till November 27, 2017 during 81 days (106 runs of observations).  The long-term light curve based on  the most dense part of  the 2015 observations  is shown in Fig. 1.  
The amplitude of the superoutburst  in the $R_C$-band was $\sim 3^m.5$. It lasted  about 18-19 days, including rebrightening and taking into account that the start of the superoutburst was around JD 2457216 according to the VSNET announcement. Note that this duration was the same as those of  another dwarf novae in the period gap NY Ser  (Pavlenko et al. 2014,  Sklyanov et al. 2018)  and MN Dra (Sklyanov et al. in preparation). The amplitude of  normal outbursts  was  a few  tenths of magnitude less and varied, which is caused by the  large variations of the  quiescent  brightness. 

\begin{figure}
\centerline{\includegraphics[width=12cm,clip=]{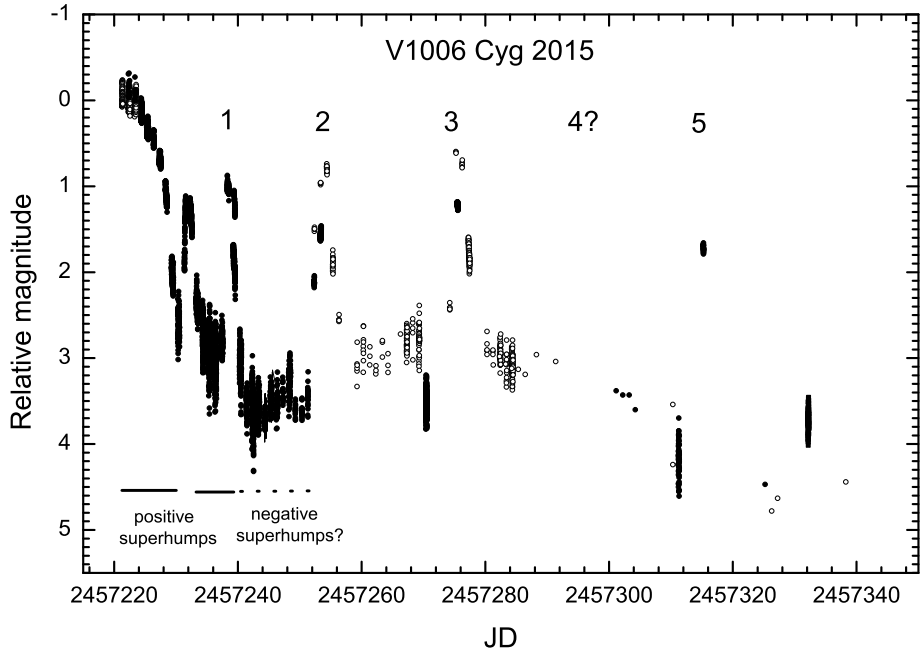}}
\vspace{-1mm}
\caption{The V1006 Cyg light curve in 2015.  $R_C$ data and unfiltered data are shown by open and filled circles, respectively.  The data are expressed relatively the comparison star No~140. 
The sequence of the normal outbursts (including a possible outburst No~4) is numbered. 
For JD 2457221 -- 2457246 the types of periodicity  are indicated.}
\vspace{-3mm}
\end{figure}

The first normal outburst occurred 6 days after the superoutburst rebrightening.  The cycle between the first and the second normal outburst lasted 16 days, while the cycle between the second and the third one 22 days. The next cycle was undefined  because of a lack of observations between JD  2457292 -- 2457301.  Potentially a normal outburst could occur in this  interval, that is too short for the superoutburst.  Taking into account that the last 2015 normal outburst  was detected in  $\sim$124 days after the start of superoutburst  and there was no possibility for the next superoutburst to occur up to  JD 2457339, we could conclude that the supercycle should be longer than 124 days. It means  that V1006 Cyg  is not an active  dwarf nova.

\subsection{Superoutburst}

We have studied the brightness variability of V1006 Cyg  during the first 25 nights since the start of the 2015 superoutburst that covered main superoutburst, rebrightening and the first normal outburst. The results are presented in Fig. 2.

\begin{figure}
	\centerline{\includegraphics[width=12cm,clip=]{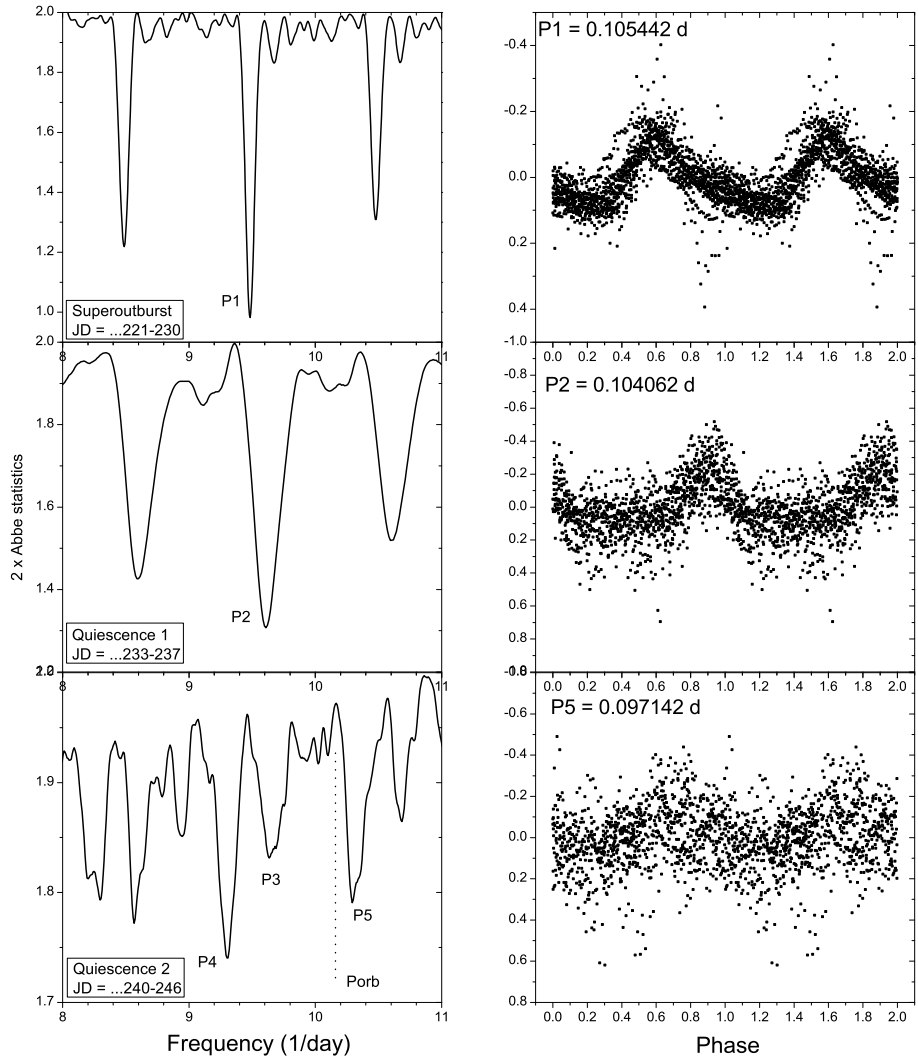}}
	\vspace{-1mm}
	\caption{Left (from top to bottom): the periodograms for the superoutburst, short-term quiescence between rebrightening and the first normal outburst and quiescence between the first and the second normal outbursts. For every data set the time interval is indicated.  The designations of peaks are given in the text. The position of the  orbital period is marked by the dotted line. The phase diagrams of corresponded data folded with the most significant periods are presented on the right.}
	\vspace{-3mm}
\end{figure}
 Positive superhumps,  with mean period $P_1=0^d.10544$(10), were detected during the superoutburst. During the short-term quiescence between the rebrightening and the start of the first normal outburst the  mean period of positive superhumps was $P_2=0^d.10406$(17). These periods are in agreement with  periods found by Kato et al. (2016) from the detailed study of the positive superhumps evolution during JD 2457219 -- 2457240.  A periodogram in quiescence after the first normal outburst points to the most significant peaks $P_4=0^d.107594$(85) and $P_5=0^d.097142$(74) that are not independent but are the day-aliased.  There is also a peak at a period $P_3$ coinciding with the period of positive superhumps $P_2$ but of lower significance,  and no indication of the orbital period. 
From two periods $P_4$ and $P_5$ we prefer the latter  because it is closer to 
a potentially possible period of negative superhumps.  

Using  the values of the orbital period 
$P_{orb}=0^d.09903$, the  positive superhump period $P_2=0^d.104062$ and the suggestive negative superhump period $P_5=0^d.09714$, one could calculate  a positive  superhump period excess  
$\epsilon^+ = (P_{+sh}- P_{orb})/P_{orb}$ and  a negative superhump period deficit  $\epsilon^- = (P_{-sh}- P_{orb})/P_{orb}$,  
where  $P_{+sh}$ and $P_{-sh}$ are the periods of positive  and  negative 
superhumps, respectively.  We obtained  $\epsilon^+ = 0.051, \epsilon^- = -0.019$ and  
$\phi = \epsilon^- / \epsilon^+ = -0.037$. 

Our data are in  good agreement with an empirical relation between $\epsilon^+, \epsilon^-$  and the orbital period (Hellier, 2001). Moreover, we found that the ratio $\epsilon^- / \epsilon^+ = -0.037$ corresponds to the mass ratio $q\sim 0.30 - 0.32$ according to the Wood's et al. (2009) model, that coincides with an independent estimate of Kato et al (2016),  $q\sim 0.26 - 0.33$. So we can conclude that our identification of the period $0^d.09714$ as the period of negative superhumps is correct. 

\subsection{Normal outbursts}

We gathered all our measured colour-indices $V-R_C$  for different stages of activity 0f V1006 Cyg  and presented them in  the color-magnitude diagram (Fig.3). All the data are attributed to the outburst decline. Unfortunately no color measurement were done at the rising part of the outbursts. Therefor, we cannot decide whether the tracks corresponding to ascending and descending branches of outbursts are the same, or they perform a loop as it was found by Smak (1978). 

\begin{figure}
\centerline{\includegraphics[width=9cm,clip=]{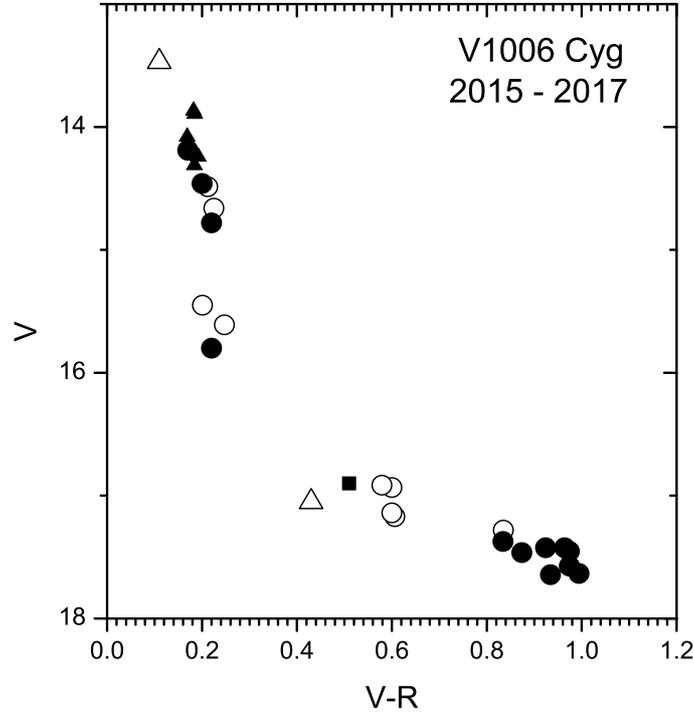}}
\vspace{-1mm}
\caption{The color-magnitude diagram for different stages of the 2015 superoutburst,  normal outbursts No~1, 2, 3 (2015) and  the wide outburst  (2017). The filled triangles, open circles, filled circles, filled triangles, filled square and  open squares  denote the mean per night data obtained with 2.6-m, 1.25-m, 70-cm, 60-cm and 50-cm telescopes respectively.}
\vspace{-3mm}
\end{figure}

The peculiarity of this color-magnitude behavior resembles those of other dwarf novae (Pavlenko et al. 2008): the small reddening during the $\sim 2^m$  brightness decline after the outburst maximum and much faster reddening during the slower approach to the quiescence  and in quiescence itself. Such behavior is probably caused  by a different contributions of the sources of radiation  to the total light  during the outburst: a decrease of accretion disk  role and increase  of the secondary component role with outburst decline.

\subsection{2017 wide outburst}

In 2017 we observed a wide outburst with the amplitude of about 3.5 mag.  The AAVSO and our data suggest its duration at least seven days (Fig.~4). During the wide outburst  decline we detected the periodic  brightness variationsin four subsequent nights. To search for the period and to compare an amplitude of periodicity  we converted detrended data of four nights  into relative intensities I${}_{rel}$ according to the formula $I_{rel} = 10^6 \cdot 10^{-0.4\cdot m}$,
where $m$ is expressed in $R_C-$ magnitudes.

\begin{figure}
\centerline{\includegraphics[width=10cm,clip=]{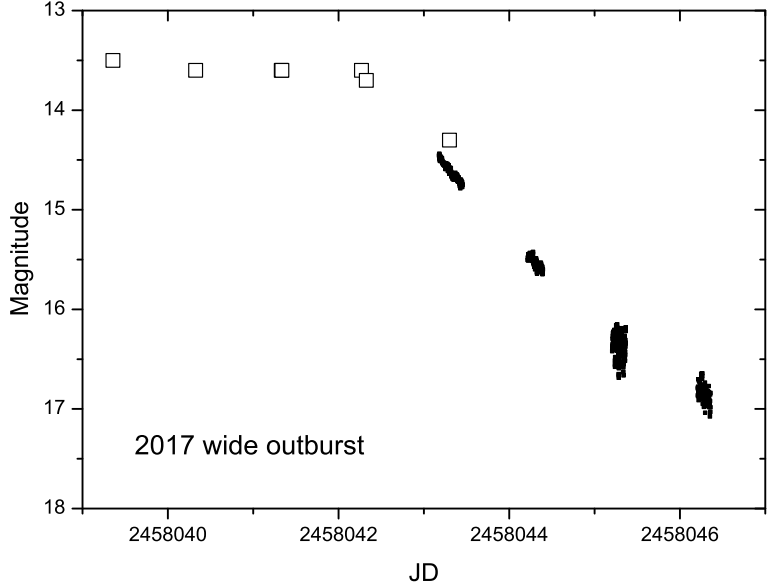}}
\vspace{-1mm}
\caption{The 2017 wide outburst. Open and solid squares denote the AAVSO (www.aavso.org) and our data respectively. }
\vspace{-3mm}
\end{figure}

The periodogram is presented in Fig. 5 (the upper panel). The most significant peak among day-aliased peaks points to a period of $0^d.09832(15)$, which coincides with the known orbital period (Pavlenko et al. 2014). Note that there were two known events of wide outbursts in 2007 and 2009 without superoutbursts, but with the orbital periodicity. 
The mean phase light curve is given in the middle panel of Fig.5. In the lower panel we show  the smoothed phase light curves for the wide outburst decline. One could see that the amplitudes of the data for the first three nights are near equal, wile the amplitude of the data of last night data  is close to quiescence is $\sim 2.5$ times lower. 

We also  detected the outburst in August 2016 during the brightness rise and the maximum in  two subsequent nights. However, a lack of further data does not allow us to specify if this was a wide or normal outburst. Since our observation at the maximum lasted about 6.5 hours, we were  able to examine  brightness variations around the orbital period, too. We constructed the periodogram for the $V$ and $R_C$ data combined together (their amplitudes are almost equal) and plotted it in the upper frame of Fig. 5. It can be seen that this periodogram also points to the periodicity around the orbital period. 

\begin{figure}
\centerline{\includegraphics[width=12cm,clip=]{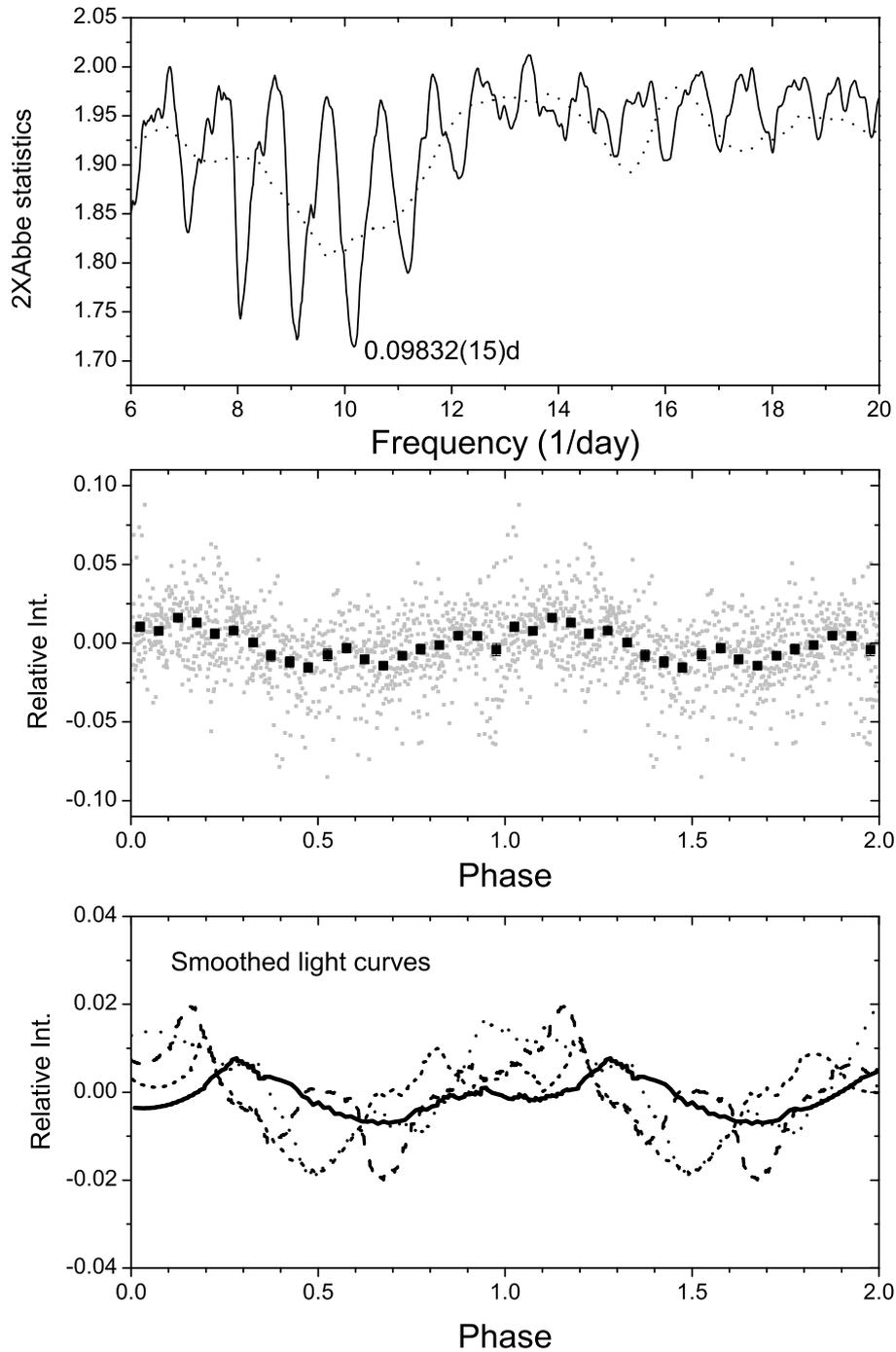}}
\vspace{-1mm}
\caption{Top: the periodogram for the data of the 2017 wide outburst decline (the solid line). The periodogram for the data of a single night of the 2016 outburst is shown by the dotted line. Middle: the mean phase light curve for the data of the wide outburst. Bottom: the original smoothed phase light curves of the wide outburst on 3 subsequent nights (dash and dotted lines). The phase light curve for the last night of the wide outburst is expressed by the solid thick line.}
\vspace{-3mm}
\end{figure}

\subsection{Quiescence}

Observations of V1006 Cyg in quiescence (out of the outbursts) were  carried out in 2015, 2016 and 2017 (see Table 1). Contrary to expectations,  we did not detect a prominent orbital modulation in this state.  This may indicate the low inclination of the binary orbit. Instead, the examples of the long runs of observations demonstrate one and the same feature of the nightly light curves   - more or less strong quasi-periodic oscillations of a variable frequency and  amplitude. The  amplitude could reach on
 some occasions $\sim 0^m.5$.  The examples of nightly light curves  and corresponding periodograms are shown in Fig.~6.  The $BVR_C$ observations for JD 2457481 showed that the amplitude of oscillations is practically the same in the $V$- and $R_C$- passband while in $B$-passband it is slightly higher.  The significance of the quasi-period depends on the amplitude of QPO and a coherence time.

For every night we  selected the more significant periods  (for definiteness the periods with significance $<1.6$). The results are given in  Table 2 and in the histogram (Fig.~7). The maximal number of quasi-periods is observed in the interval 20-30 minutes that is 5-7 times shorter than the orbital period. The quasi-periods between 50-60 min probably are doubles of those between 20-30 min.  Note that the potential orbital modulation is probably  hidden by  QPOs. 

It is possible to find similar high-amplitude QPOs in CVs that have no relation to their orbital periods. The behavior and nature of QPOs  at various states of  cataclysmic variables activity may be different (see e.g.: Kato et al. 2017b; Shugarov et al. 2016; Pavlenko and Shugarov  1999; Scaringi et al. 2017; Pavlenko 1996; Zemko et al. 2014). A detailed 
study of QPOs in  CVs in quiescence during quiescence is limited by the faintness of these objects. 

Our results are in  good agreement with the model of compactions in an accretion disk described by Fridman and Bisikalo (2008). According to their model, there could appear a coagulate in an accretion disk after the binary underwent a decrease of  its accretion rate. This coagulate moves with a period  of $\sim 0.15P_{orb} - 0.18P_{orb}$ and could be understood as the one-armed spiral density wave (Fridman et al. 2003; Fridman and Khoruzhii 2003). 

\begin{figure}
\centerline{\includegraphics[width=12cm,clip=]{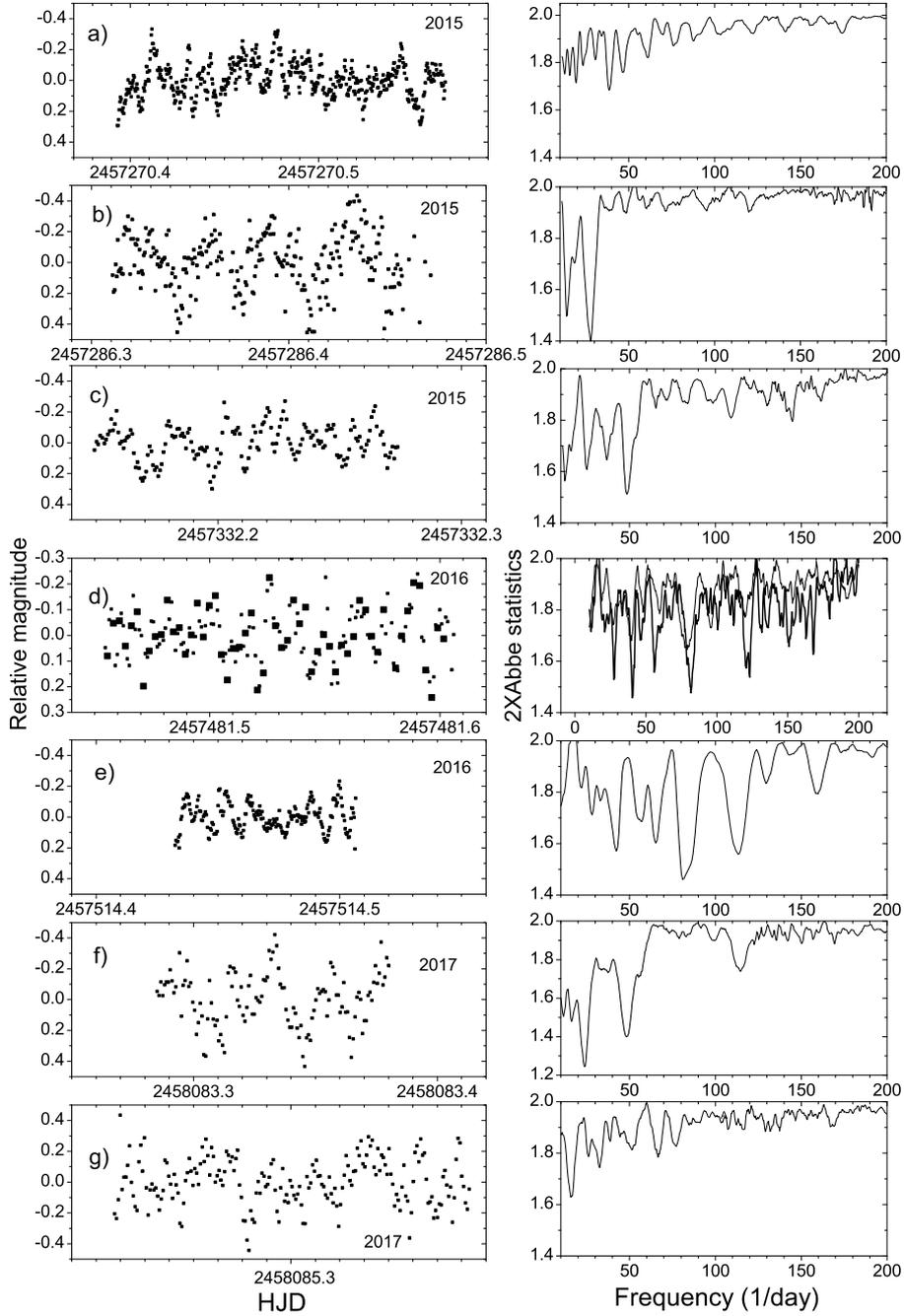}}
\vspace{-1mm}
\caption{Left panels: the examples of nightly light curves of V1006 Cyg during quiescent states in 2015, 2016 and 2017 normalized to the mean brightness level.  The light curve at JD 2457481 contains both $VR_C$ and $B$ data  plotted by dots and bold squares, respectively. Right panels: corresponding periodograms. There are two periodograms for the JD 2457481 on the same graph: the periodograms for $V$  $R_C$ and $B$ data are expressed by the thin and bold lines, respectively.}
\vspace{-3mm}
\end{figure}

\begin{table}[t]
\small
\begin{center}
\caption{The most significant quasi-periods for selected nights of 2015, 2016 and 2017.}
%\label{ts}
\begin{tabular}{ll}
\hline\hline
JD &QPO (min) \\
\hline
2457286	  & 53; 96\\
2457332	  & 30; 115 \\
2457481	  & 18; 26; 36; 51\\
2457514	  & 18; 23; 33 \\
2458083	  & 29; 63; 90; 120 \\
\hline\hline
\end{tabular}
\end{center}
\end{table}

\begin{figure}
\centerline{\includegraphics[width=10cm,clip=]{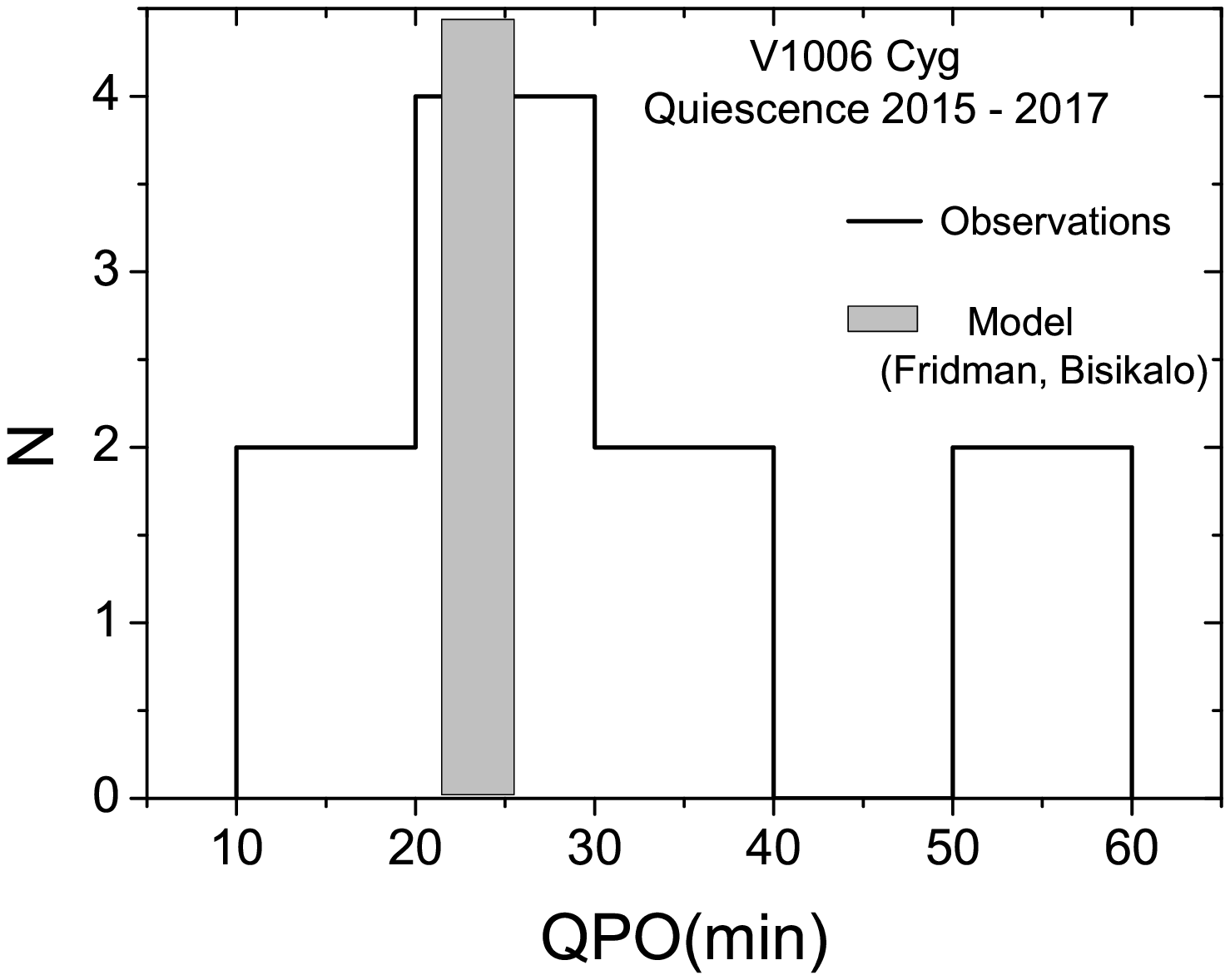}}
\vspace{-1mm}
\caption{The histogram of the formally most significant QPOs observed in quiescences. The gray region 
corresponds to a range of predicted QPOs according to the Fridman and Bisikalo (2008) model.}
\vspace{-3mm}
\end{figure}

\section{Conclusion}

The detailed study of the SU UMa-type nova-in the gap, V1006 Cyg,  allowed us 
to find the peculiarities of its superoutburst, normal and wide outbursts, 
and quiescence. We define its activity state as a supercycle that is longer than 124 d and variable cycles that are 16 d and 22 d for the 2015 year.  
In 2017 we observed a wide outburst similar to those observed in 2007 and 2009  
(Pavlenko et al. 2014) that was accompanied, as in the two previous cases, 
by the orbital brightness variations without superhumps. The orbital signal at the start 
of the wide outburst decline was 2.5 times higher than those at the end. If the orbital 
modulation is caused by the hot spot visibility over the orbital period, this could mean 
a slightly enhanced mass transfer rate. The absence of superoutbursts means a hard achievement 
of the tidal resonance in every wide outburst for a binary with the critical mass ratio of 
0.26--0.33, which is close to the stability limit of the tidal instability (Kato et al. 2016). 
We found  positive superhumps during the 2015 superoutburst that lasted  up to the first and probably the second normal outburst. The strongest signal in quiescence between the first two normal outbursts could be  a negative superhump. The ratio of the  negative superhump period deficit to the positive superhump period excess corresponds to the mass ratio $q = 0.30 - 0.32$ according 
to the Wood at al. (2009) model and coincides with  $q\sim 0.26-0.32$  found by Kato et al. (2016). The orbital signal in quiescence may be  masked  by the high-amplitude QPOs  caused by a probably low inclination of the orbit.  The values of the V1006 Cyg quiescent QPOs are in  good agreement with Fridman and Bisikalo  (2008) prediction  of  heterogeneities in the accretion disk caused by the mass accretion rate decrease.

\acknowledgements
The authors thank an anonymous referee for valuable comments and suggestions.
This work was partially funded by RFBR according to the research project 18-32-00371, by the subsidy allocated to Kazan Federal University for the state assignment in the sphere of scientific activities (3.9780.2017/8.9); the Russian Government Program of Competitive Growth of Kazan Federal University and was performed with the use of observational data obtained at the North-Caucasus Astronomical Station of KFU; S. Shugarov~thanks for partial support the grants VEGA 2/0008/17 and APVV 15-0458. S. Simon is grateful to V.P. Lapchuk for technical help in observations and to N.A. Beisen for the hospitality and provision of the observational time at the ``Bobek'' observatory.

\end{document}